# Semiconductor-based superlens for sub-wavelength resolution below the diffraction limit at extreme ultraviolet frequencies

**M.A. Vincenti[1,2], A. D'Orazio[2], M.G. Cappeddu[1,3], Neset Akozbek[1], M.J. Bloemer[1], M. Scalora[1]**

[1]*Charles M. Bowden Research Center, AMSRD-AMR-WS-ST, RDECOM, Redstone Arsenal, Alabama 35898-5000, USA; email: vincenti@deemail.poliba.it*

[2]*Dipartimento di Elettrotecnica ed Elettronica, Politecnico di Bari, Via Orabona 4, 70125 Bari, Italy*

[3]*Dipartimento dei Materiali, Università di Roma "La Sapienza", Via Eudossiana 18, 00184 Rome, Italy*

We theoretically demonstrate negative refraction and sub-wavelength resolution below the diffraction limit in the UV and extreme UV ranges using semiconductors. The metal-like response of typical semiconductors such as GaAs or GaP makes it possible to achieve negative refraction and super-guiding in resonant semiconductor/dielectric multilayer stacks, similar to what has been demonstrated in metallo-dielectric photonic band gap structures. The exploitation of this basic property in semiconductors raises the possibility of new, yet-untapped applications in the UV and soft x-ray ranges.

PACS numbers: 42.25.Bs; 42.25.Fx; 42.70.Qs; 42.79.Ag; 78.66.-w;

## I. INTRODUCTION

In recent years there has been an increased interest in a super-resolving lens based on negative index materials (NIMs), which are not found in nature. NIMs have been fabricated in the microwave regime by manipulating the effective electric permittivity and magnetic permeability using metallic inclusions. Although the same approach is now being used to push towards the optical regime, the effort has proven to be difficult. Some limitations can be overcome by using a material with a negative permittivity (i.e. metal) and TM-polarized incident light. It has been demonstrated that light propagating inside a silver layer can refract negatively, and the layer behaves like a superlens in the ultraviolet (UV) range [1-3]. This peculiar behavior is achieved when the operating wavelength is tuned below the plasma frequency, where the permittivity of transition metals achieves negative values. However, the performance of a simple metal layer is frustrated by the opacity that characterizes transition metals. An attempt to overcome the low transmittance of single metal layers was made by introducing systems based on non-resonant metallo-dielectric (MD) multilayers [4-7]. Although those systems are able to resolve object below the wavelength limit, the structures were still characterized by relatively low transmittance. In order to improve the transparency of transition metals below and far from the plasma frequency, resonant metallo-dielectric photonic band gap (MD-PBG) configurations were been proposed [8-12], which consist of alternating metallic and dielectric layers having thicknesses of the order of the incident wavelength, so that resonance tunneling and field localization inside both metal and dielectric layers can occur. The advantage of this configuration is that the transparency region can be tuned over a desired spectral range with a transparency of 50% or more.

Propagation effects have been recently discussed in realistic, resonant MD-PBGs in the context of negative refraction and super-resolution in the visible and near infrared ranges [13], and it has been shown that the resonance tunneling regime contributes to a waveguiding phenomenon that confines the light along the transverse coordinate within a small fraction of a wavelength, practically suppressing diffraction. For instance, an aperture 12nm wide could be imaged with 600nm-wavelength nearly 350nm away, resulting in a spot size of approximately 60nm. This effect occurs as a result of negative refraction of the Poynting vector inside each metal layer, balanced by normal refraction inside the adjacent dielectric layer: the degree of field localization and material dispersion together determine the total momentum that exists in any layer, and thus the direction of energy flow.

It was recently demonstrated that a semiconductor-based metamaterial exhibits negative refraction in the far infrared ($\lambda$~10μm) [14], where absorption losses are relatively small. However, most semiconductors are also characterized by a negative permittivity in the deep UV wavelength range. For example, the dielectric constant of GaAs, shown in Fig.1, is negative between approximately 120 and 270nm. Other semiconductors like GaP and Ge behave similarly, with slightly different upper and lower bounds. Compared to bulk metals, the negative permittivity region of semiconductors can reach much shorter wavelengths, and thus may find important applications in extreme UV (XUV) lithography or molecular spectroscopy. Lithography has evolved significantly over the years starting from the visible (g-line 436) towards short wavelengths in order to make smaller feature sizes. Current state of the art lithography uses 197nm sources. However, next-generation lithography is being developed at 157nm and even short wavelengths at 135nm. For example, Intel Corporation is developing new XUV lithography technologies to be able to print circuit features at a 32nm scale [15]. Currently, the smallest feature size achievable is 45nm using the combination of a 193nm source and immersion liquids [16]. Semiconductors have been widely studied in photonics, and are perhaps the most ubiquitous materials used in the electronic chip industry. As such, semiconductors are unique and can be doped to allow the engineering of its effective electromagnetic properties. Therefore, it is entirely possible to design the same kind of resonant superlenses illustrated in references [11-13] that work at XUV wavelengths and in the soft x-ray range.

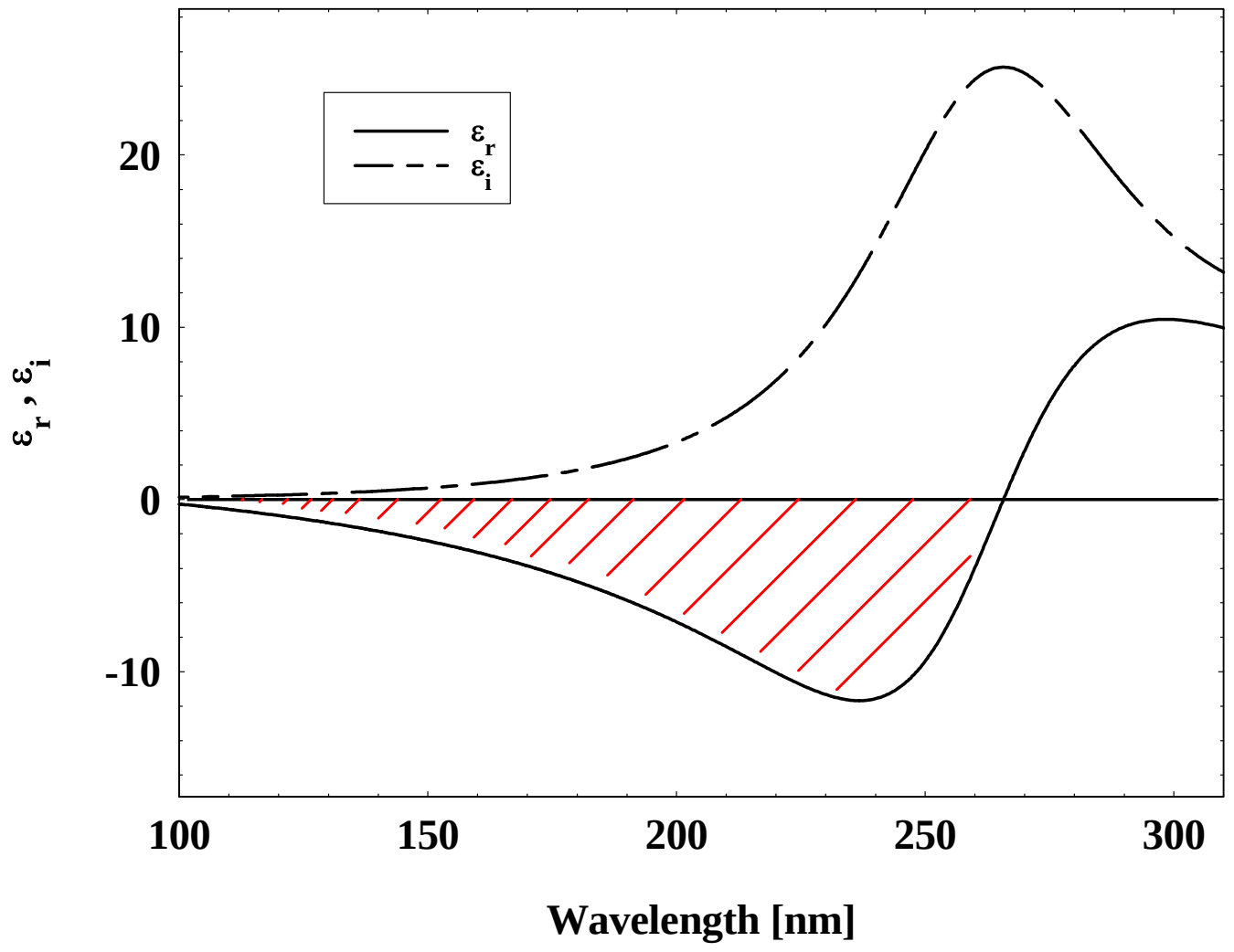

Fig.1: Real and imaginary parts of the dielectric permittivity of GaAs below 300nm.

## II. THE SEMICONDUCTOR-BASED LENS

The balance of the total electromagnetic momentum that distinguishes the MD-PBG multilayer stacks, which limits diffraction, can also be achieved in semiconductor/dielectric multilayer stacks at wavelengths that are ordinarily thought to be inaccessible. As an example we consider the stack sketched in Fig.2, which is composed of three 14nm GaAs layers and 16-nm-thick layers of dielectric material *X*, coated by two anti-reflection coating layers of material *X* (8nm thick). This arrangement is identical in nature to the symmetric arrangement of resonant metallo-dielectric layers discussed in references [12, 13]. Material *X* is assumed to be loss-less with dielectric permittivity $\varepsilon$=6.28. This choice of dielectric permittivity is required to balance the negative refraction process that occurs inside the semiconductor layers. The plane-wave transmittance of the system that includes the material *X* is calculated using the standard transfer matrix method: in Fig.3-a the transmission and reflection spectra are plotted for a TM-polarized wave incident at 45° with respect to the propagation direction.

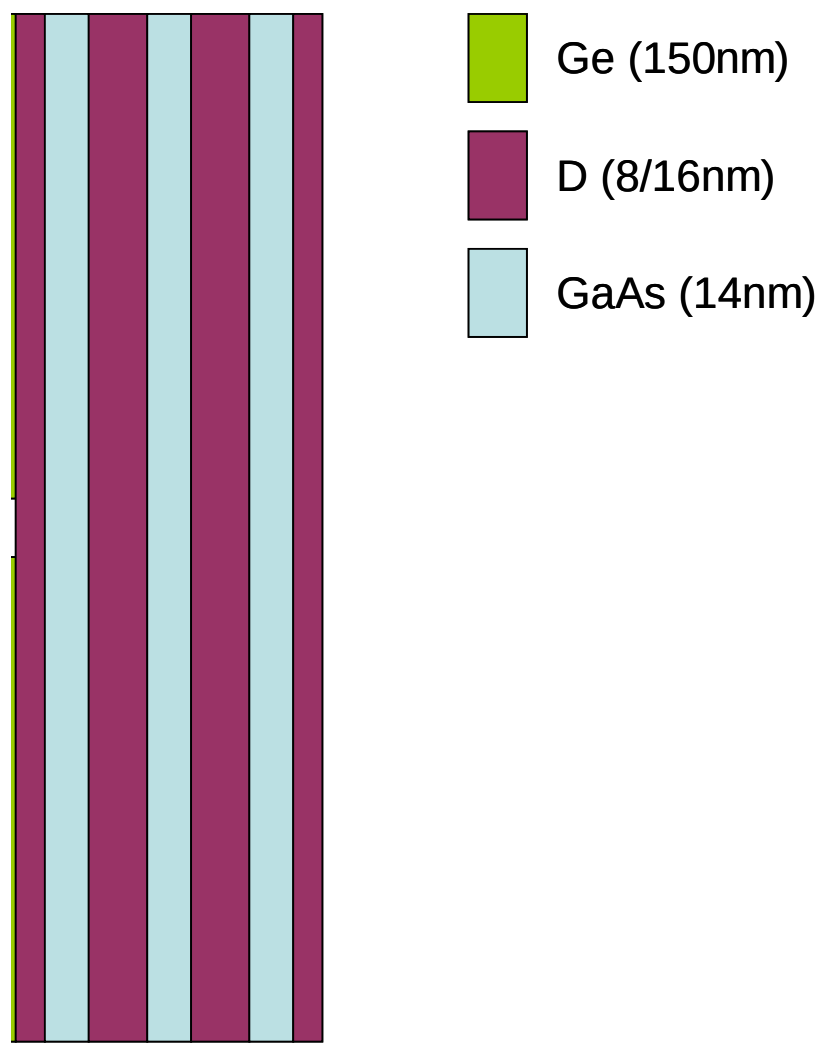

Fig.2: Sketch of the multilayer.

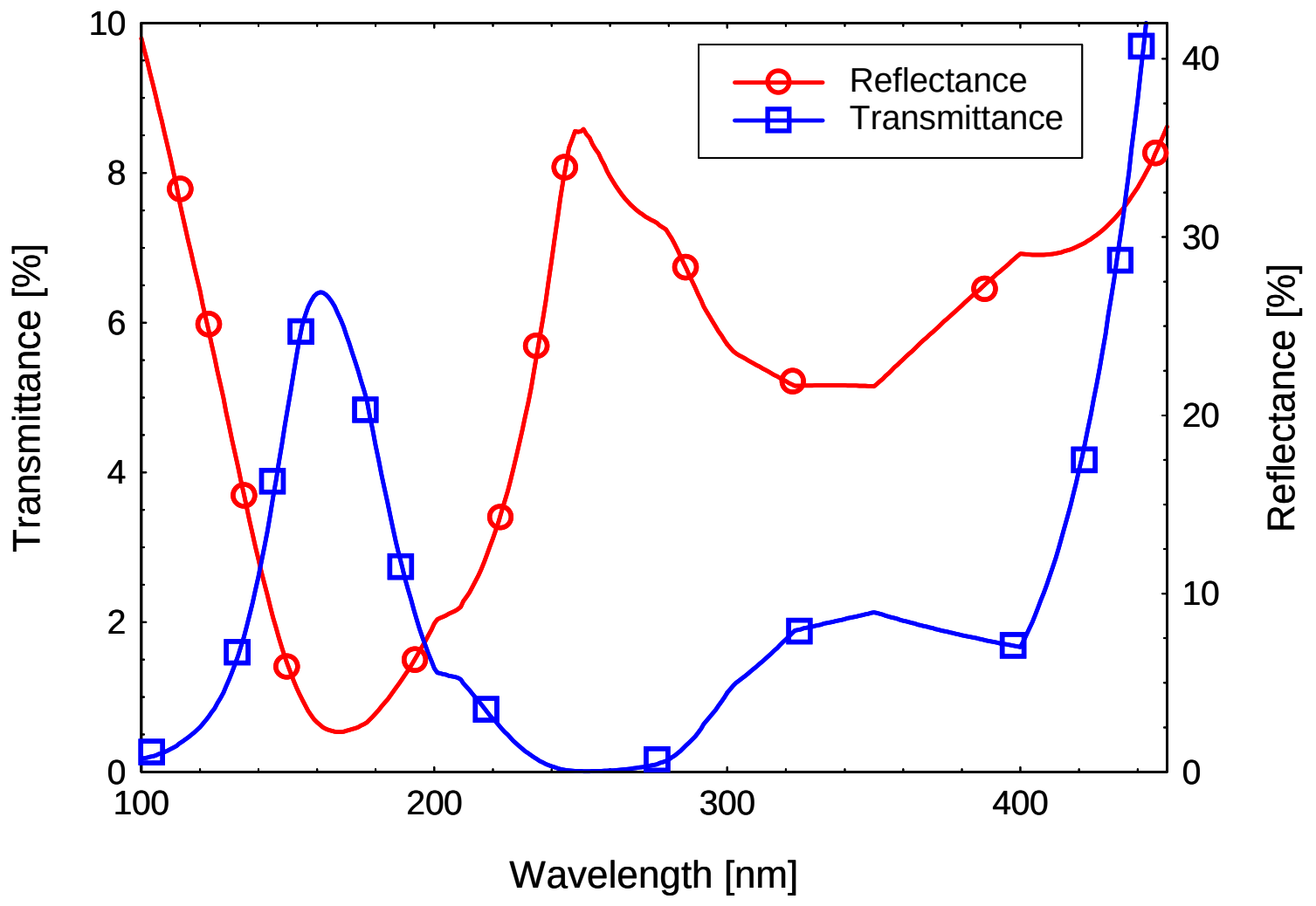

Fig.3-a: Transmittance and Reflectance of the 3.5 period GaAs(14nm)/*X*(16nm) structure of Fig.2 without the Germanium substrate. Incidence is at 45º.

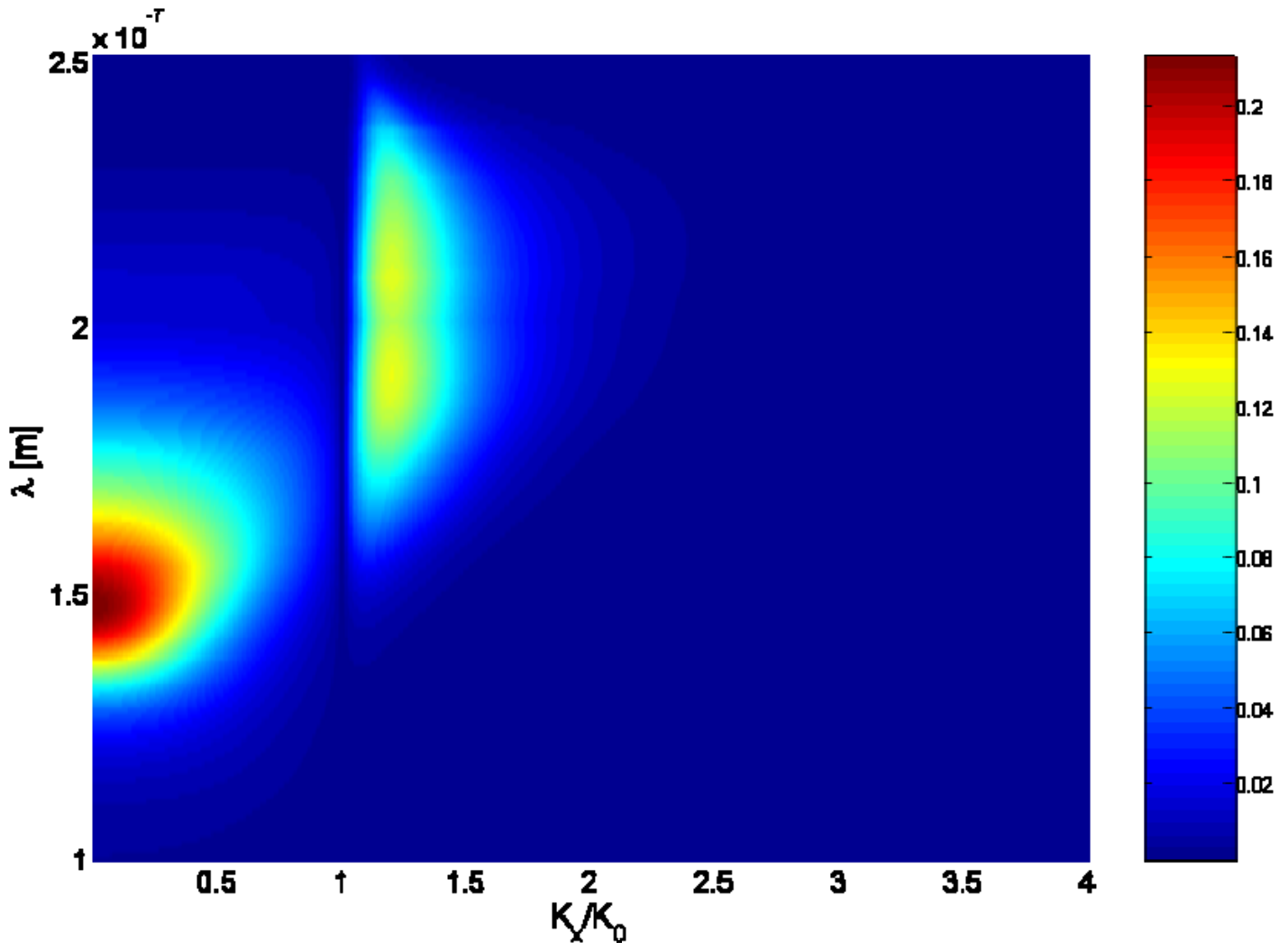

Fig.3-b: Dispersion relation for the stack of Fig.2 without the Germanium substrate.

The optical transfer function $T(k_x/k_0,\lambda)$ of this particular stack (Fig.3-b), also calculated by means of the standard transfer matrix method [13], reveals the behavior of this semiconductor-based lens for the whole range of wave-vector components: the observable value of transmittance in the propagation mode region ($k_x<k_0$) reaches around 20% at normal incidence; at the same time $T(k_x/k_0,\lambda)$ also provides information about the coupling mechanism of evanescent wave-vector components, or guided modes ($k_x>k_0$), with the multilayer structure.

In order to reproduce the light emitted by a small object having subwavelength features, we consider a rectangular-shaped slit in an opaque dielectric substrate [11,13] 150nm thick that absorbs and reflects the wave completely outside the slit region. The calculations and results were performed and obtained in a two-dimensional simulation grid using a finite-difference, time-domain (FDTD) method, and a Fast Fourier Transform (FFT) based pulse propagation algorithm. The grid was discretized as follows: $\Delta x=\Delta z=2$ nm for both the transverse ($x$ axis) and longitudinal ($z$ axis) coordinates. The dielectric constant of GaAs [17] at 172nm is ε=-1.997+i3.582. While the stack does not exhibit high transmittance and low reflectance simultaneously [13] in the wavelength range under investigation (see Fig.3-a), the transmittance of a single 45nm GaAs layer at 172nm is only ~0.5%. This should be compared to the ~6%, or 10 times larger, achieved by the resonant multilayer under investigation. We demonstrate that this stack frustrates the diffraction process thanks to negative refraction that occurs in each of the GaAs layers, which effectively replaces the metal layers in resonant MD-PBG structures [12,13]. In spite of the large absorption coefficient in GaAs, evanescent modes are still able to couple inside the stack resulting in efficient confinement of the incident field (see Fig.3-b).

Choosing an operating wavelength to control diffraction inside the lens is not trivial, and cannot generally be done by examining the $T(k_x/k_0,\lambda)$ alone: unlike propagating modes, evanescent modes cannot be decoupled from the source. Instead, an excellent measure of whether super-resolution can be achieved is given by the direction of the total electromagnetic momentum inside the multilayer stack (Fig.4): if the total momentum of the wave that traverses the stack is mostly longitudinal with a small-enough transverse velocity component, then a quasi-stationary surface wave having velocities of order $10^{-4}$c is excited (at 172nm), and evanescent modes are transported across the stack [13].

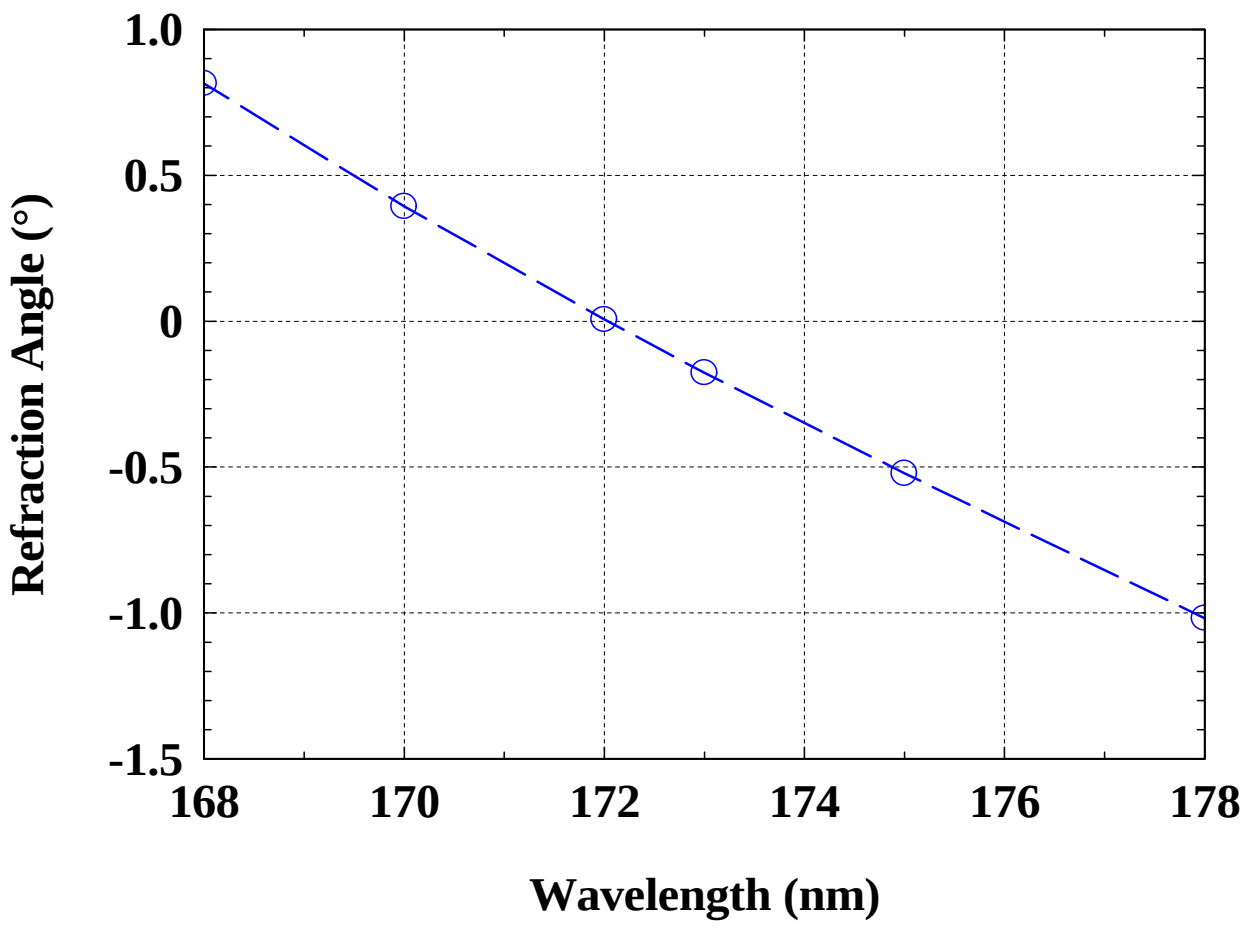

Fig.4: Refraction angle for a wave packet incident at 45 degrees. It is zero at 172nm, which indicates a mostly longitudinal momentum, super-guiding and excitation and transport of evanescent modes.

Moreover, our lens cannot be considered by means of an effective medium approach because of its field localization properties (Fig.5) and cannot be treated as a perfect lens by any means, unlike non-resonant metallo-dielectric stacks proposed by other authors [4,5,7,18-20]. While in the effective medium approach both super-guiding and super-resolution arise as a result of the large anisotropy induced by the alternation of positive and negative dielectric constants under non-resonant conditions, the structure we propose is characterized by field localization effects. The localization properties of the *E* and *H* fields are substantially different: the *E* field is more intense inside the dielectric layers, while the *H* field is more intense inside the semiconductor layers, just as occurs in metallo-dielectric resonant structures [13].

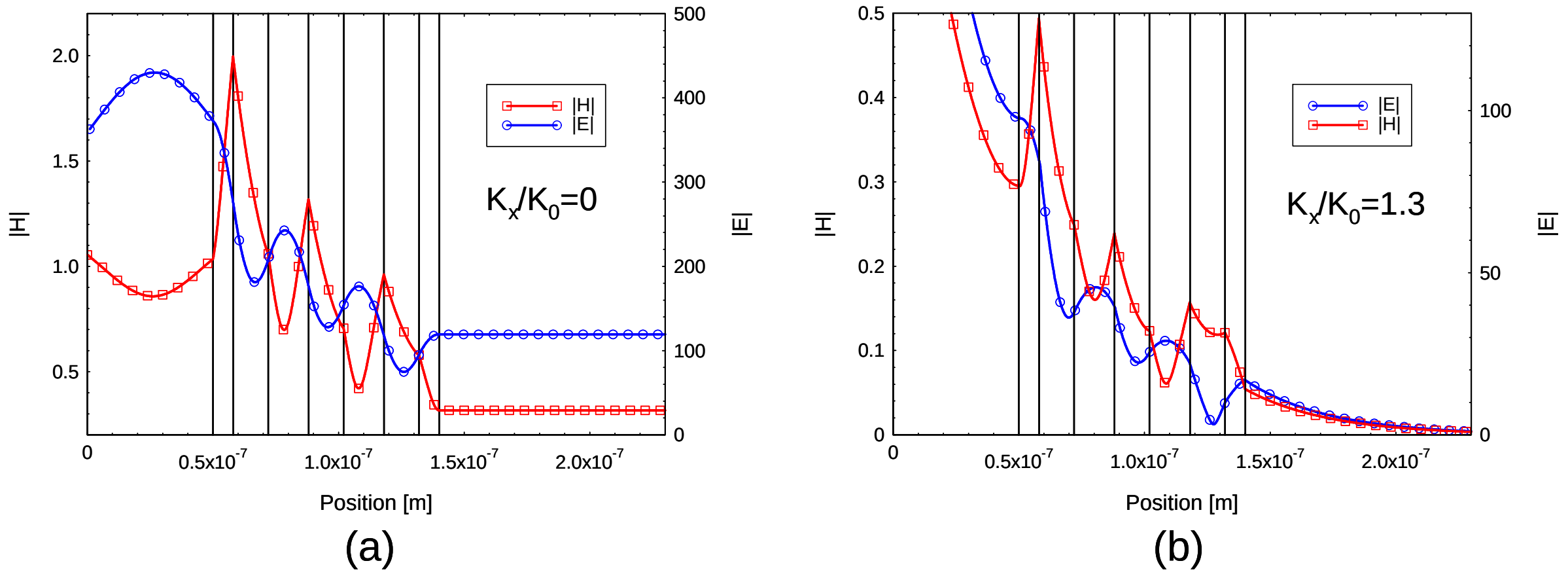

Fig.5: Electric and magnetic field amplitudes inside the stack for the mode a) $kx/k0$=0 and b) $kx$ /$k0$=1.3. Inside the stack, evanescent fields are quite similar to propagating modes. Outside the stack evanescent modes rapidly decay.

To verify the super-guiding and super-resolving ability of the lens we also examined the full-width at half-maximum (FWHM) of the electric and magnetic fields, as well as the Poynting vector at the exit interface of the lens: for a 40nm wide slit the FWHM of the longitudinal Poynting vector at the exit interface of the stack is about 54nm, about a factor of three smaller compared to the FWHM of the free-space propagating Poynting vector (154nm) and of a beam which propagates through 90nm of dielectric *X* (130nm). Therefore, diffraction is suppressed and the device acts as a super-waveguide for a slit ~λ/4 wide (see Fig.6).

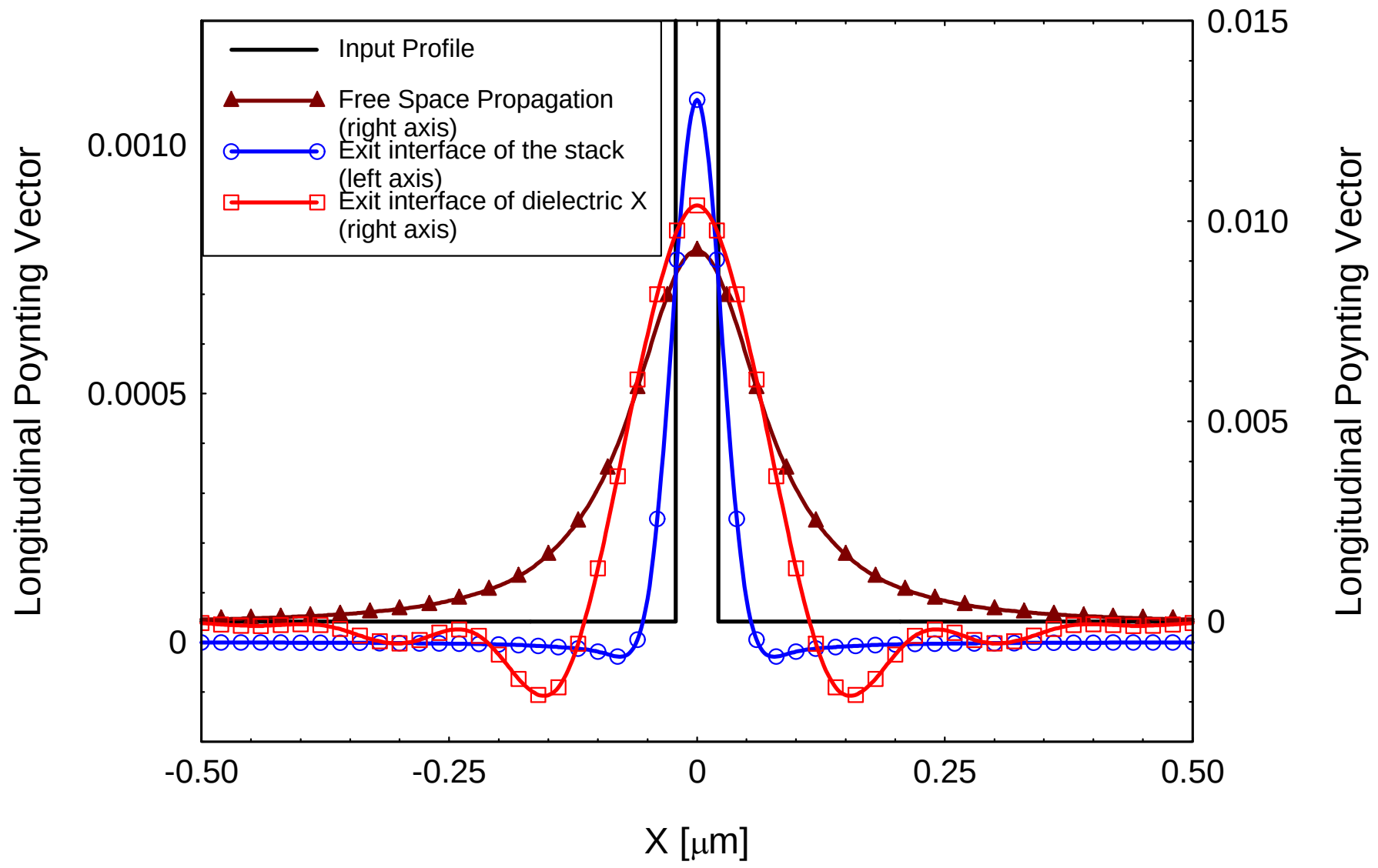

Fig.6: Longitudinal Poynting vector profiles at the exit interface with (blue curve) and without (red curve) the lens. The brown curve is the longitudinal Poynting vector profile at the exit interface of 90nm (equivalent to the thickness of the lens) of dielectric *X*. The black rectangle represents the source profile.

In order to quantify the resolving power of the lens we consider the same structure depicted in Fig.2, in which two slits 40nm wide have been carved on the opaque, absorbing substrate. Our calculations suggest that our stack yields a resolution of approximately 20% (Rayleigh criterion) when the center-to-center distance between the slits is s~82nm. Just as was the case for the single-slit system, here too we predict significant improvement with respect to the free-space propagation and propagation through 90nm of dielectric *X* (equivalent to the lens thickness) of the same sources (see Fig.7).

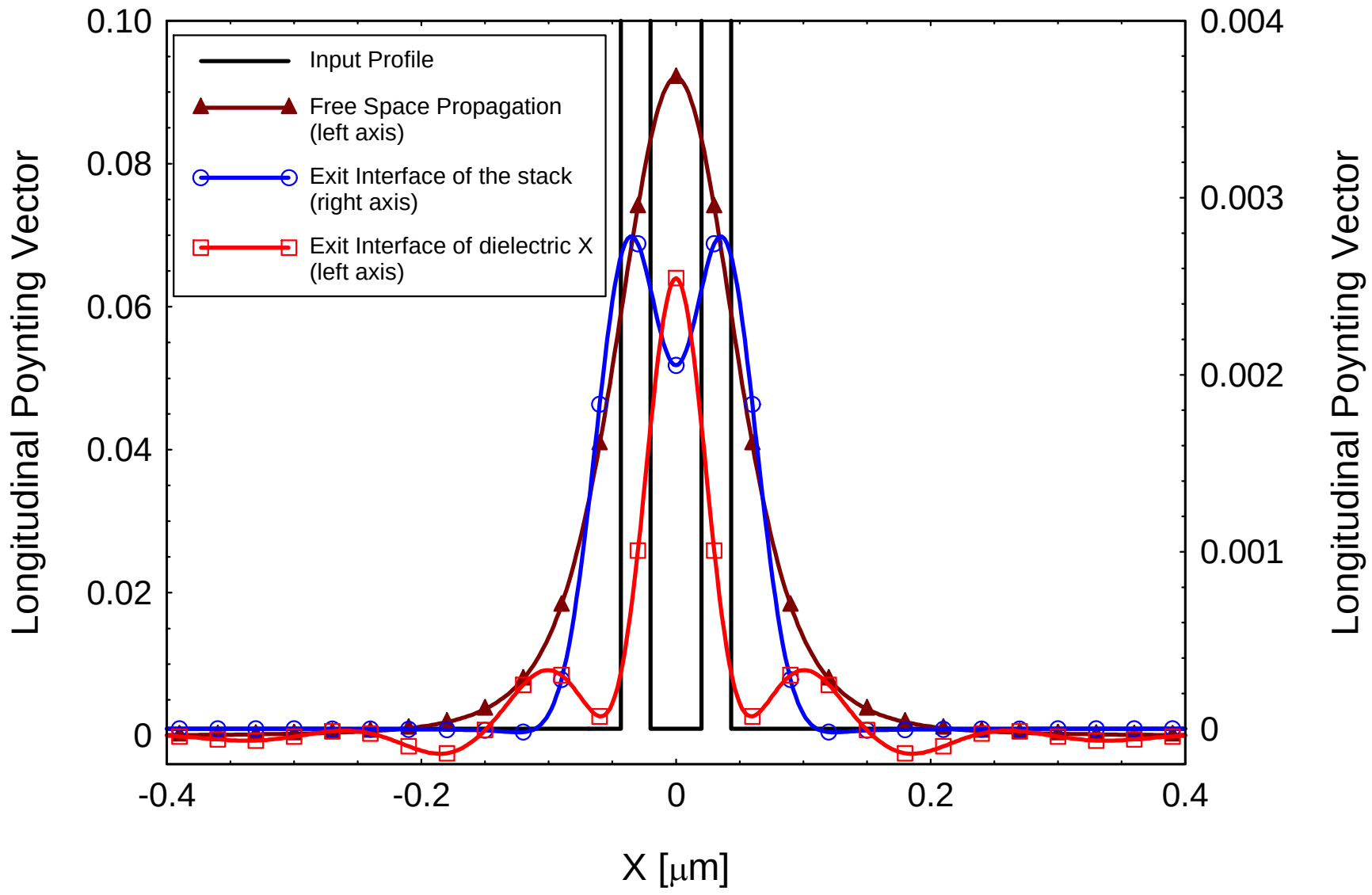

Fig.7: Longitudinal Poynting vector profiles at the exit interface of the two slits system with (blue curve) and without (red curve) the lens. The brown curve is the longitudinal Poynting vector profile at the exit interface of 90nm (equivalent to the thickness of the lens) of dielectric *X* for two 40nm wide slits 82nm apart. The black rectangle represents the source profile.

## III. FEASIBLE LENSES

As mentioned in the previous paragraph, a realistic semiconductor-based lens also needs an appropriate dielectric material that meets certain conditions. For example, materials like MgO, KCl and $Si_3N_4$ are all good candidates because they display relatively small absorption in the wavelength range of interest. But this short list by no means exhausts all the possibilities. Here we discuss two realistic semiconductor/dielectric structures composed of GaAs/MgO and GaAs/KCl. For simplicity we consider the stack sketched in Fig.2 in which the layers of dielectric material *X* are in turn replaced with MgO and KCl. The dispersion profiles of the dielectric materials are found in Palik's handbook [17]. The plane-wave transmittance of both systems calculated using the standard transfer matrix method is reported in Fig.8: the relatively absorption of MgO and KCl influences the transparency of the stack only marginally (compare Figs. 8 and 2).

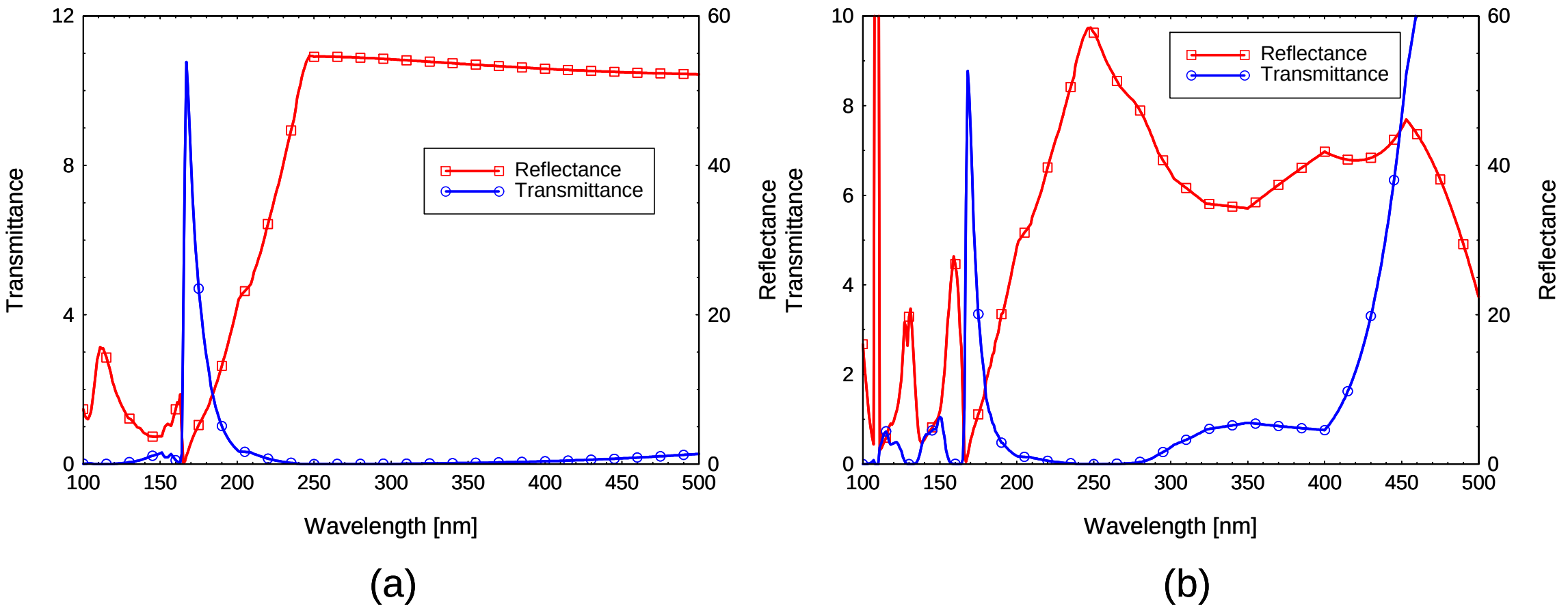

Fig.8: Transmission and reflection spectra of the a) 3.5 period GaAs(14nm)/MgO(16nm) and b) 3.5 period GaAs(14nm)/KCl(16nm) structures for a TM-polarized wave at normal incidence.

We also calculated the optical transfer function $T(k_x/k_0,\lambda)$ (Fig.9) of the two realistic stacks. The function $T(k_x/k_0,\lambda)$ essentially defines the system's effective numerical aperture (NA), and determines how the lens handles propagating and evanescent modes. Both these semiconductor/dielectric lenses behave similarly to an ideal lens (i.e. one where the dielectric material is absorptionless) for the whole range of wave-vector components. $T(k_x/k_0,\lambda)$ also sheds light on the possible ways to improve the value of the NA for a specified operating wavelength by properly engineering the multilayer structure. Moreover, the large absorption coefficient of GaAs and the relatively small absorption introduced by the dielectrics still allows the coupling of evanescent modes inside the stack and provides for an efficient confinement of the incident field.

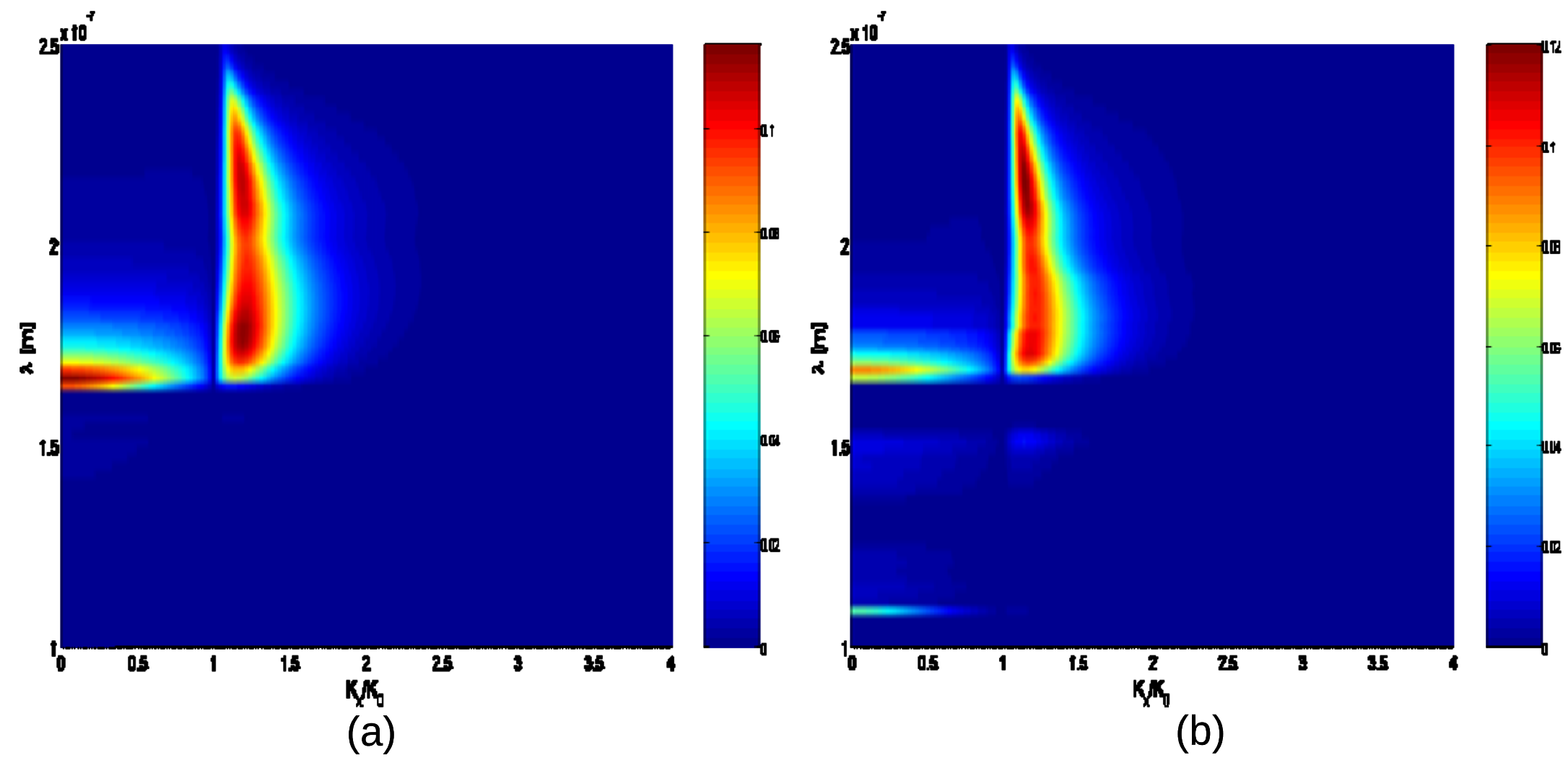

Fig.9: Optical transfer function of the a) 3.5 period GaAs(14nm)/MgO(16nm) and b) 3.5 period GaAs(14nm)/KCl(16nm) structures.

In order to test the resolving power of the GaAs/MgO and GaAs/KCl realistic lenses that we have discussed we considered two rectangular-shaped slits carved on the same opaque dielectric substrate described above, and an input source tuned at λ=172nm. As shown in Fig.10, the diffraction process for two 20nm wide slits (~λ/8) is significantly reduced when the center-to-center distances between the slits are 80nm, 70nm and 66nm, well below ~λ/2: the KCl-based system (Fig.10-a) reaches visibilities of 65%, 31% and 19% respectively, while the MgO-based multilayer stack yields visibilities values of 57%, 25% and 15% .

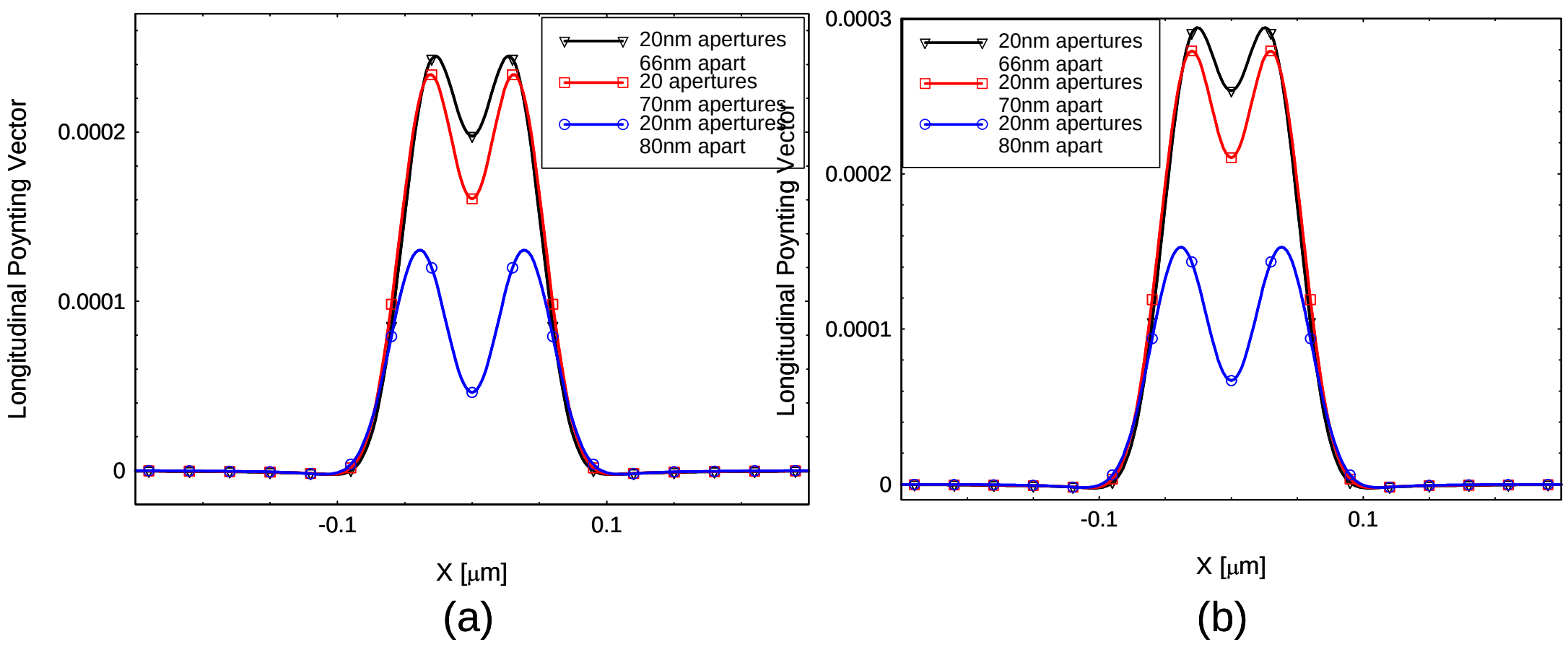

Fig.10: Longitudinal Poynting vector profiles at the exit interface of the two 20nm wide slits system when the dielectric *X* is replaced by (a) KCl and (b) MgO.

## IV. CONCLUSIONS

We have shown that negative refraction of the Poynting vector occurs for TM-polarized waves inside a semiconductor/dielectric multilayer stack for wavelengths where the dielectric constant of the semiconductor becomes negative. The nanostructure is capable of super-guiding and sub-wavelength resolution below the diffraction limit despite the relatively large absorption characteristic of the UV resonance. We discussed two realistic semiconductor-based lens configurations, demonstrating that material losses do not alter substantially the peculiarities of such systems. Our findings open the way to new approaches to negative refraction of light in semiconductors and new applications, particularly for plasmonic devices and new optical components in XUV lithography systems, in regimes that are usually thought to be impossible to access.

## REFERENCES

[1] J. B. Pendry, Phys. Rev. Lett. **85**, 3966 (2000).

[2] D. O. S. Melville, and R. J. Blaikie, Opt. Express **13**, 2127 (2005).

[3] N. Fang, H. Lee, C. Sun, and C. X. Zhang, Science **308**, 534 (2005).

[4] Z. Liu, H. Lee, Y. Xiong, C. Sun, X. Zhang, Science **315**, 1686 (2007).

[5] S. A. Ramakrishna et al., J. Mod. Opt. **50**, 1419 (2003).

[6] B. Wood, J. B. Pendry, D. P. Tsai, Phys. Rev. B **74**, 115116 (2006).

[7] P. A. Belov and Y. Hao, Phys. Rev. B **73**, 113110, (2006).

[8] M. Scalora et al., J. Appl. Phys. **83**, 2377 (1998).

[9] M. J. Bloemer and M. Scalora, Appl. Phys. Lett. **72**, 1676 (1998).

[10] M. Scalora et al., Opt. Photon. News **10**, 23 (1999).

[11] M. Scalora et al., Opt. Express **15**, 508 (2007).

[12] M. J. Bloemer et al., Appl. Phys. Lett. **90**, 174113 (2007).

[13] D. de Ceglia et al., Phys. Rev. A **77**, 033848 (2008).

[14] A. J. Hoffman et al., Nature Mater. **6**, 946 (2007).

[15] B. Wu and A. Kumar, J. Vac. Sci. Technol. B **25,** 1743 (2007).

[16] M. Rothshild et al., Lincoln Lab. Jour. **14,** 221 (2003).

[17] E. D. Palik, *Handbook of Optical Constants of Solids* (Academic Press, New York 1985).

[18] K. J. Webb and M. Young, Opt. Lett. **31**, 2130 (2006).

[19] X. Li, S. He and Y. Jin , Phys. Rev. B **75**, 045103 (2007).

[20] A. Salandrino and N. Engheta, Phys. Rev. B **74**, 075103 (2006).